\documentclass[prl,reprint,amssymb]{revtex4-2}

\usepackage{xcolor}

\usepackage{amsmath}  
\usepackage{amsfonts}
\usepackage{graphicx}
\usepackage{hyperref}
\hypersetup{
    colorlinks=true,
	citecolor=blue,
    linkcolor=blue,
	urlcolor = blue
}

\begin{document}

\title{Anisotropic skyrmion mass induced by surrounding conduction electrons: A Schwinger-Keldysh field theory approach}

\author{Felipe Reyes-Osorio}
\author{Branislav K. Nikoli\'c}
\email{bnikolic@udel.edu}
\affiliation{Department of Physics and Astronomy, University of Delaware, Newark, DE 19716, USA}

\date{\today}

\begin{abstract}
The current-driven motion of magnetic skyrmions, as topologically protected winding vector fields of local magnetization, has attracted considerable attention due to both fundamental interest in the dynamics of topological solitons, and potential spintronic applications. While widely-used Thiele equation with zero skyrmion mass captures its rigid motion under the assumption of preserved skyrmion shape, experiments on realistic skyrmions find deviation due to inertial mass with numerous attempts to account for it by invoking internal skyrmion dynamics, geometry of the nanowire, and interaction with magnons or phonons. However, experimentally observed skyrmion mass is often much larger than predicted by these theories. In this Letter, we employ Schwinger-Keldysh field-theoretical approach to study conduction electrons interacting with localized magnetic moments comprising a skyrmion to obtain a stochastic differential equation of its motion. By analyzing its time-retarded damping kernel, we extract an \textit{anisotropic} mass governed by conduction electrons and demonstrate its substantial effect on the current-driven skyrmion motion, even in the absence of any skyrmion deformation.
\end{abstract}

\maketitle

\begin{figure}[t]
    \centering
    \includegraphics[width = \linewidth]{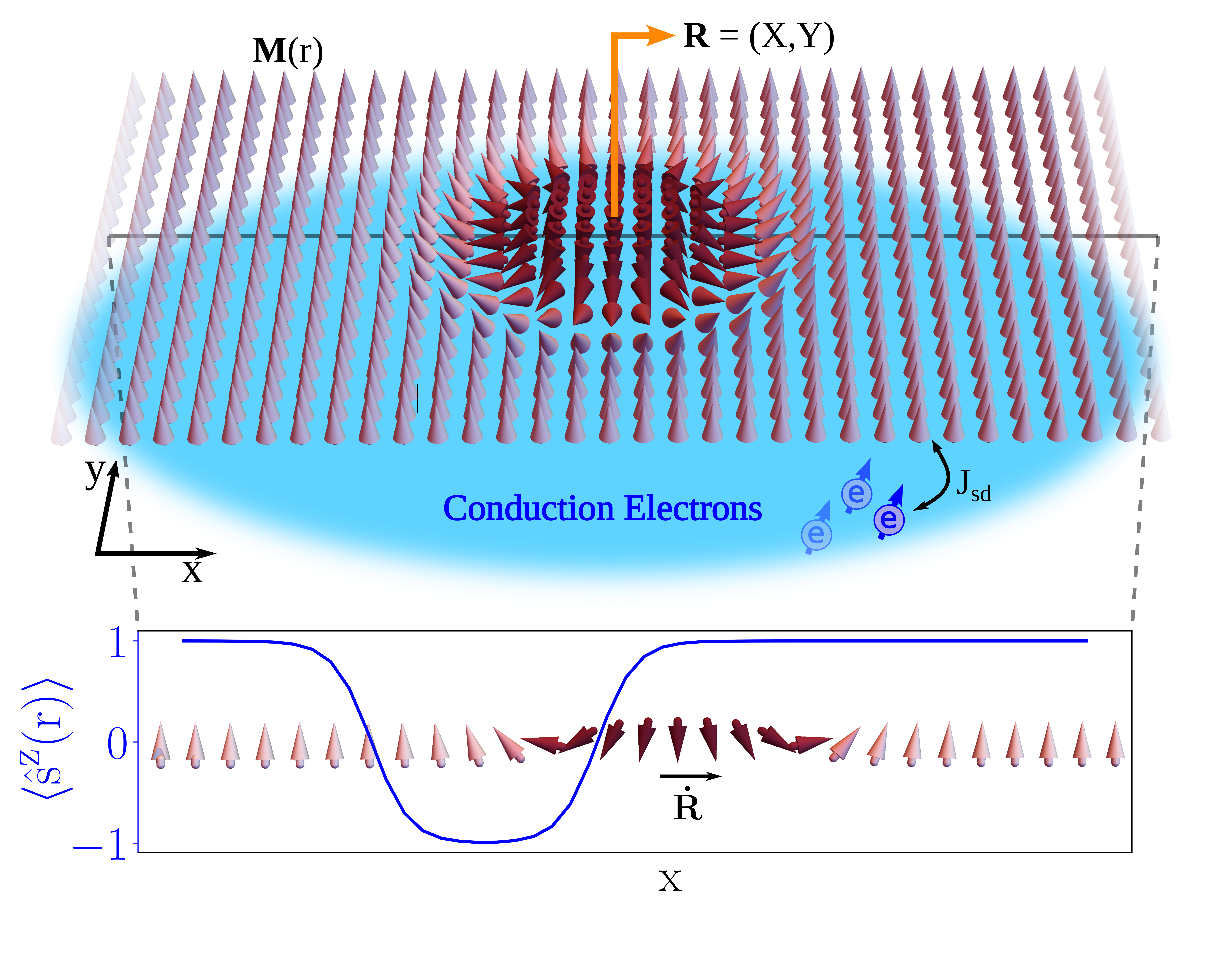}
    \caption{Schematic view of a single skyrmion as a 2D texture comprised of noncolinear and winding localized magnetic moments (red arrows), with topological charge $Q=1$ [Eq.~\eqref{eqQ}], interacting with a bath of conduction electron spins (blue arrow) within ferromagnetic metal, via \textit{sd} exchange $J_{sd}$. The bottom panel illustrates how the skyrmion moves through the static spatial profile of the electronic spin density $\langle \hat{\mathbf{s}}\rangle (\mathbf{r})$ where the misalignment between this vector and $\mathbf{M}(\mathbf{r})$ leads to local spin torque~\cite{Ralph2008, Petrovic2018} $\propto \langle \hat{\mathbf{s}}\rangle (\mathbf{r}) \times \mathbf{M}(\mathbf{r})$ whose integral generates mass term in the Thiele equation [Eq.~\eqref{eqFinalEOM}].}
    \label{figSetup}
\end{figure}

\textit{Introduction.}---The dynamics of topological solitons in magnetism~\cite{Braun2012} has attracted considerable attention due to both fundamental interest, allowing access to phenomena~\cite{Manton2004} in cosmology and high-energy physics using table-top experiments~\cite{Woo2017, Petrovic2021}, and applications for memory and logic devices~\cite{Koshibae2015, Vakili2021}, including beyond von Neumann~\cite{Raab2022} and quantum computing~\cite{Psaroudaki2021}. In particular, skyrmions \cite{Nagaosa2013, Goebel2021, Dohi2022}, winding textures of the local magnetization, have attracted considerable attention due to their topological protection against scattering and thermal fluctuations, as well as the advent of heterostructures of ultrathin transition-metal ferromagnets and heavy metals with strong spin-orbit coupling hosting skyrmions at room temperature~\cite{Boulle2016, Dohi2022}, or more recent discovery of two-dimensional (2D) magnets and their van der Waals heterostructures offering additional knobs to generate and control skyrmions~\cite{Park2021, Sun2021}. 

Since many applications rely on precise control of skyrmion motion by spin-polarized currents or external fields, one of the key tasks for the theory has been the derivation of the proper equation of motion in terms of a small set of collective coordinates. For example, Thiele~\cite{Thiele1973} derived a massless equation of motion
\begin{equation}\label{eqFinalEOM}
    \underbrace{m\dot{\mathbf{v}}_d}_{\text{mass term}}+ \underbrace{G(\mathbf{v}_s -\mathbf{v}_d) + D(\beta\mathbf{v}_s - \alpha\mathbf{v}_d) = -\nabla U}_{\text{massless Thiele equation}},
\end{equation}
for the skyrmion center $\mathbf{R}=(X, Y)$ [Fig.~\ref{figSetup}] in the case of its rigid (i.e., without skyrmion shape deformation) and steady-state translation. Here $\mathbf{v}_d$ is the drift velocity of the skyrmion center, $\mathbf{v}_s$ is the velocity of the polarized electrons, $G_{ij} = \epsilon_{ij}\tilde Q$ describes the Magnus force, $D$ is the dissipative coefficient, $\beta$ measures the strength the nonadiabatic spin-transfer torque (STT)~\cite{Tatara2019}, $\alpha$ is the Gilbert damping in the Landau-Lifshitz-Gilbert (LLG) theory of the dynamics of localized magnetic moments (LMMs) $\mathbf{M}(\mathbf{r})$ viewed as classical vectors, and $-\nabla U$ is a conservative force acting on the soliton, such as a trapping potential $U(\mathbf{r})$. 

However, experiments \cite{Buettner2015}, and micromagnetic numerical simulations \cite{Ivanov2005, Sheka2006, Moutafis2009, Schuette2014, Komineas2015} point to the need to add a mass term in Eq.~\eqref{eqFinalEOM}. This need is typically traced~\cite{Makhfudz2012, Wu2022, Suzuki2022} to the change in the shape of the skyrmion and the corresponding energy cost, particularly in transient situations. In these explanations, the mass term [Eq.~\eqref{eqFinalEOM}] is obtained solely from the LLG classical theory. Nevertheless, the experimentally observed mass~\cite{Buettner2015} is often orders of magnitude larger than such theoretical estimates. This discrepancy has prompted the inclusion of additional effects, such as interaction with magnons~\cite{Psaroudaki2018} or phonons~\cite{Capic2020}, as well as the effect of the curvature of the nanowire~\cite{Pavlis2020}. On the other hand, interaction with conduction electrons, which must be described quantum-mechanically and whose currents are most frequently employed to drive skyrmion motion, remains largely unexplored. An exception is Ref.~\cite{Martinez2017} where mass due to electrons was added, but only as phenomenological term. 

\begin{figure}
    \centering
    \includegraphics[width=\linewidth]{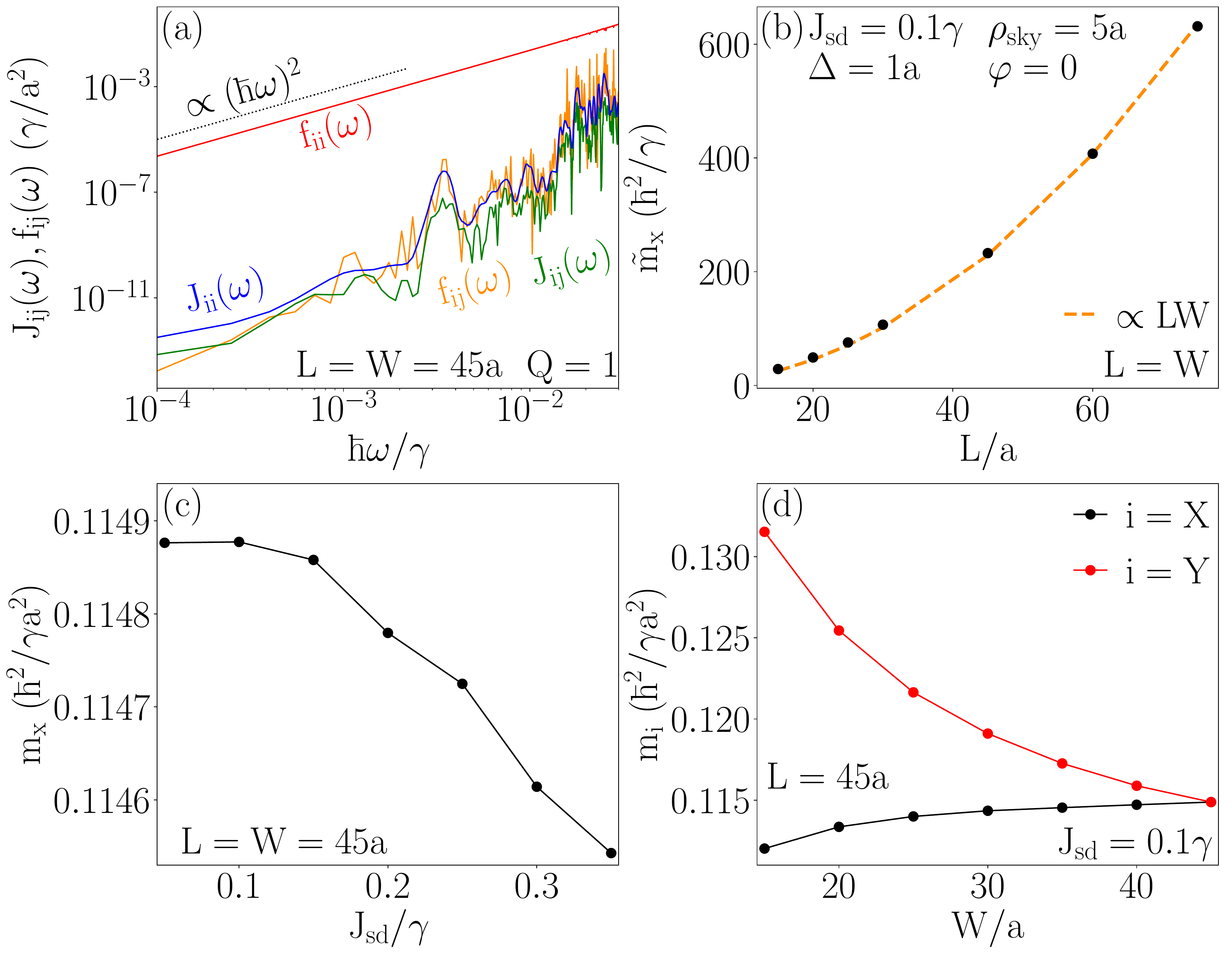}
    \caption{(a) Diagonal ($i=j$) and off-diagonal ($i\neq j$) elements of the spectral $J_{ij}(\omega)$ and dissipative $f_{ij}(\omega)$ components of the damping kernel $\eta_{ij}(\omega)$ [Eqs.~\eqref{eqFinalAction} and~\eqref{eq:spectralDissipative}]. The only relevant component at low frequencies is $f_{ii}(\omega)$, which gives rise to skyrmion mass plotted as a function of relevant parameters $J_{sd}$, $L$, and $W$ in (b)--(d). The mass $\tilde m_i$ in panel (b) contains the true mass $m_i= \tilde m_i/LW$.  In all panels we use $\mu=1.5\gamma$ and $T=300$~K.}
    \label{figMass}
\end{figure}

In this Letter, we derive microscopically---that is by starting from Hamiltonians of LMMs and conduction electrons and using the Schwinger-Keldysh functional integral~\cite{Hurst2020, Kamenev2011}---the mass term for the Thiele Eq.~\eqref{eqFinalAction}. Our mass term turns out to be \textit{highly anisotropic}, that is it depends on the longitudinal and transverse direction with respect to the length of the nanowire. We note that the electronically-induced mass unveiled here is nonzero, even for perfectly rigid skyrmion motion~\cite{Wu2022, Makhfudz2012}. Physically, it arises due to the inability of the electron spin $\langle \hat{\mathbf{s}}\rangle (\mathbf{r})$ to track the LMMs $\mathbf{M}(\mathbf{r})$ of a skyrmion, except in adiabatic limit~\cite{Bajpai2020, Stahl2017} of unrealistically large \textit{sd} interaction, $J_{sd} \rightarrow \infty$ [Fig.~\ref{figMass}(c)]. This always leads to an additional STT~\cite{Ralph2008, Petrovic2018} $\propto \langle \hat{\mathbf{s}}\rangle (\mathbf{r}) \times \mathbf{M}(\mathbf{r})$ on the LMMs~\cite{Petrovic2021, Bajpai2020, Sayad2015}.

\textit{Schwinger-Keldysh formalism for noncolinear magnetic textures interacting with electrons.}---In order to study the non-equilibrium dynamics of the system, we formulate the action $S = S_M + S_e$ along the so-called Keldysh closed contour $\mathcal{C}$ \cite{Kamenev2011}. The action is divided into two parts
\begin{subequations}
\begin {eqnarray}
S_M &=& \int_\mathcal{C}\! dt d\mathbf{r} \, \Big\{\dot{\mathbf{M}}(\mathbf{r},t)\cdot\mathbf{A} - \mathcal{H}[\mathbf{M}(\mathbf{r},t)]\Big\}, \label{eqMagAction} \\
S_e &=& \int_\mathcal{C}\! dt d\mathbf{r} \, \bar\psi(\mathbf{r},t) \Big\{ i\partial_t - H[\mathbf{M}(\mathbf{r},t)] \Big\}\psi(\mathbf{r},t), \label{eqElecAction}
\end{eqnarray}
\end{subequations}
where $S_M$ describes the LMMs, while $S_e$ describes the conduction electrons. Here, $\dot{\mathbf{M}}\cdot\mathbf{A}$ is the Berry phase term~\cite{Nagaosa2006, Stone1996} with $\dot{ \mathbf{M}}\equiv \partial\mathbf{M}/\partial t$; $\psi = (\psi_\uparrow, \psi_\downarrow)^T$ is a Grassmann spinor of the electron field~\cite{Kamenev2011}; $\mathbf{r}=(x,y)$ is the position vector in 2D and $t$ is time. The quantum Hamiltonian $H$ of conduction electrons is given by (using $\hbar=1$)
\begin{equation}
    H[\mathbf{M}(\mathbf{r},t)] = -\frac{1}{2m_e}\nabla^2 - J_{sd} \mathbf{M} (\mathbf r,t)\cdot \boldsymbol{\sigma},
\end{equation}
where $m_e$ is the mass of the electron, $\boldsymbol{\sigma}=(\hat\sigma_x, \hat\sigma_y, \hat\sigma_z)$ is the vector of the Pauli matrices, and the second term of $H$ represents the coupling between the time-dependent LMMs and the spin of conduction electrons.

Considering the rigid propagation of a skyrmion, we track its geometric center~\cite{Wu2022} using collective coordinates $\mathbf{R} = (X, Y)^T$, which are promoted to dynamical variables. Since the energy of the skyrmion is independent of its location, we can treat $\mathcal{H}[\mathbf{M}]$ as constant, and focus only on the Berry phase contribution \cite{Stone1996}. By decomposing all quantities defined on the contour $\mathcal{C}$ into their forward ($+$) and backwards ($-$) components, this term can be interpreted as the difference between the areas bounded by the forward and backward trajectories of the magnetization on the unit sphere~\cite{Altland2010}. Furthermore, we apply the Keldysh rotation for a real field \cite{Kamenev2011} by changing the variables
\begin{equation}
    \mathbf{R}^\pm = \mathbf{R}_c \pm \frac{1}{2}\delta \mathbf{R},
\end{equation}
where $\mathbf{R}_c$ describes the classical evolution and $\delta\mathbf{R}$ corresponds to quantum fluctuations. Note that while the notation $\delta\mathbf{R}$ is standard for the quantum component of evolution, it is not necessarily small. Nevertheless, the stochastic equations of motion derived in Eq.~\eqref{eqRawEOM} are obtained in the semi-classical limit where such quantum fluctuations enter only as a small random force (i.e., noise). Therefore, we will assume small $\delta\mathbf{R}$, $|\delta\mathbf{R}| \ll |\mathbf{R}_c|$. Under these conditions, a procedure formally equivalent to that of Ref. \cite{Stone1996} yields the action
\begin{equation}
S = 4\pi Q \int\! dt \, \delta R^i \epsilon_{ij} \dot R_c^j
\end{equation}
where $(R^{1}, R^{2}) \equiv (X,Y)$; $\epsilon_{ij}$ is the 2D Levi-Civita symbol; repeated indices are summed over; and $Q$ is the topological charge of a skyrmion defined by $Q = \frac{1}{4\pi} \int\! d\mathbf{r} \, \mathbf{M} \cdot \partial_x \mathbf{M} \times \partial_y \mathbf{M}$. This charge represents the number of times the local magnetization, modelled as a classical vector field $\mathbf{M}(\mathbf{r})$, maps onto a sphere~\cite{Nagaosa2013}.

In the electronic action $S_e$, we use $\psi = \psi\big(\mathbf{r}-\mathbf{R}(t), t\big)$ for the electron field assumed to adjust to the skyrmion motion, so that the instantaneous Hamiltonian reduces to the static one $H=H(\mathbf{M}_0)$, as illustrated in the bottom panel of Fig.~\ref{figSetup}. Thus, the skyrmion affects the conducton electrons only through its velocity originating from the time derivative in Eq.~\eqref{eqElecAction}, so that the electron action becomes
\begin{equation}\label{eq:electronActionVel}
    S_e = \int\! dt d\mathbf{r} \, \Big\{\bar\psi \left( i\partial_t - H[\mathbf{M}_0] \right)\psi - i \bar\psi \dot{\mathbf{R}}\cdot\nabla\psi\Big\}.
\end{equation}
The next step is to transform to the eigenbasis of the electron Hamitonian assuming static skyrmion, $\psi = \sum_{\mathbf k\sigma} \phi_{\mathbf k\sigma}(r) c_{\mathbf k\sigma}$, where $c_{\mathbf k\sigma}$ are the new Grassmann spinors and $\phi_{\mathbf k\sigma}$ are the single-particle wavefunctions satisfying $H\phi_{\mathbf k\sigma} = \varepsilon_{\mathbf k\sigma}\phi_{\mathbf k\sigma}$. Here $\mathbf k$ is the wavevector, and $\sigma$ is polarization~\cite{Littlejohn1991} rather than spin, denoting whether the electron state (anti)aligns with the underlying magnetic texture, thus belonging to the (upper)lower spin-split band. Using this eigenbasis, we obtain the following action
\begin{equation}
    S_e = \int\! dt \, \Big\{ \bar c_{n} \left( i\partial_t - \varepsilon_{n} \right)c_{n} - \dot R^i V^i_{nn^\prime} \bar c_{n} c_{n^\prime} \Big\}, \label{eqElecActionDiag}
\end{equation}
where $n$ abbreviates the quantum numbers $\mathbf k,\sigma$, and  $V^i_{nn^\prime} = i\int dr \ \phi_n^* \partial_i \phi_{n^\prime}$ are the matrix elements of the momentum operator $(\hat p_x, \hat p_y) \equiv (-i\partial_1, -i\partial_2)$ in the eigenbasis. Applying the Keldysh rotation to the electrons~\cite{Kamenev2011}, $c^\pm = \frac{1}{\sqrt 2}(c_1 \pm c_2)$ and $\bar c^\pm = \frac{1}{\sqrt 2}(\bar c_2 \pm \bar c_1)$, yields
\begin{equation}
    S_e = \int \! dtdt^\prime \, \Big\{ \bar{\mathbf{c}}_n \check G^{-1}_n \mathbf{c}_n - V^i_{nn^\prime} \bar{\mathbf{c}}_n \check R^i \mathbf{c}_{n^\prime} \Big\}, \label{eqElecActionKeldysh}
\end{equation}
where $\mathbf{c}=(c_1, c_2)^T$, and $\check O$ denotes a matrix in the Keldysh space~\cite{Kamenev2011}
\begin{equation}
    \check G_n(t, t^\prime) = \begin{pmatrix}
        G^R_n & G^K_n \\
        0     & G^A_n
    \end{pmatrix},
    \quad \check R^i = \begin{pmatrix}
        \dot R^i_c  & \delta\dot R^i/2 \\
        \delta\dot R^i/2    & \dot R^i_c 
    \end{pmatrix}.
\end{equation}
Here $G^{R/A/K}_n$ are the retarded/advanced/Keldysh Green's functions (GFs)~\cite{Kamenev2011} of conduction electrons.

The electrons in Eq. \eqref{eqElecActionKeldysh} can be integrated out to second order in the skyrmion velocity $\dot R_c$ as a small parameter. The total action $S=S_M + S_e$ then becomes
\begin{eqnarray}\label{eqFinalAction}
    S &=& 4\pi Q \int\! dt \, \delta R^i\left(\epsilon_{ij} \dot R_c^j - F^i \right) \nonumber \\
    &-& \int\! dt dt^\prime \, \eta_{ij}(t,t^\prime)\delta R^i(t) R^j_c(t^\prime)  \\ 
    &+& \frac{i}{2}\int\! dtdt^\prime \, \ C_{ij}(t,t^\prime) \delta R^i(t) \delta R^j(t^\prime), \nonumber
    \end{eqnarray}
where $F^i$ is the nonequilibrium force exerted by the electrons onto the skyrmion;  $\eta_{ij}(t,t^\prime)$ is the nonlocal-in-time (i.e., time-retarded) damping kernel, akin to the ones derived microscopically~\cite{Bajpai2018, Sayad2015, Hurst2020, Nunez2008} in other problems where quantum electrons interact with classical LMMs or postulated phenomenologically~\cite{Thonig2015, Bose2011}; and $C_{ij}(t,t^\prime)$ determines the correlation between quantum fluctuations. The latter two quantities are given in terms of the polarization functions~\cite{Hurst2020}
\begin{subequations}
\begin{eqnarray}
    \Pi^R_{ij} &=& \sum_{nn^\prime} V^i_{nn^\prime}V^j_{n^\prime n}\big[G^R_n(t,t^\prime)G^K_{n^\prime}(t^\prime, t) \\ 
    &+& G^K_n(t,t^\prime)G^A_{n^\prime}(t^\prime, t) \big] \nonumber \\
    \Pi^K_{ij} &=& \sum_{nn^\prime} V^i_{nn^\prime}V^j_{n^\prime n}\big[G^R_n(t,t^\prime)G^A_{n^\prime}(t^\prime, t) \\
    &+& G^A_n(t,t^\prime)G^R_{n^\prime}(t^\prime, t) +  G^K_n(t,t^\prime)G^K_{n^\prime}\big] \nonumber.
\end{eqnarray}
\end{subequations}
Here $\eta_{ij} = \frac{i}{2}\partial_t\Pi^R_{ij}$ and $C_{ij}=\frac{1}{4}\partial_t\partial_{t^\prime} \Pi^K_{ij}$, where time derivatives are taken while neglecting any implicit time dependence of the non-equilibrium electronic distribution function, i.e., effectively taking an adiabatic approximation. The third term in Eq.~\eqref{eqFinalAction} that is quadratic in the quantum fluctuations can be replaced through the Hubbard-Stratonovich transformation~\cite{Altland2010, Hurst2020} with a classical Gaussian noise $\xi^i(t)$ which is correlated according to $C_{ij}(t, t^\prime)=\langle \xi^i(t)\xi^j(t^\prime)\rangle$. Therefore, by minimizing the action with respect to the quantum fluctuations, we finally obtain the Langevin-type~\cite{Farias2010} stochastic equations of motion for a skyrmion coupled to conduction electrons
\begin{equation}\label{eqRawEOM}
    \epsilon_{ij}\tilde Q \dot R^j + \int\! dt^\prime \, \eta_{ij}(t, t^\prime) R^j(t^\prime) = F^i + \xi^i, 
\end{equation}
where $\tilde Q = 4\pi Q$ and we drop the subscript $R_c^i \mapsto R^i$ for simplicity of notation. The first term in Eq.~\eqref{eqRawEOM} is the Magnus force~\cite{Wu2022, Stone1996} that governs the usual massless Thiele dynamics, which is \textit{purely classical}, while other terms are \textit{quantum corrections} due to conduction electrons. We also note that our derivation leading to Eq.~\ref{eqRawEOM}, is similar to the one applied to electron-induced mass of a ferromagnetic domain wall (DW)~\cite{Hurst2020}, where in both cases noncolinear magnetic textures are affected by conduction electrons through the matrix elements $V^i_{nn^\prime}$. Thus, we conjecture that the Schwinger-Keldysh theory used here and in Ref.~\cite{Hurst2020} can be applied to any noncolinear magnetic texture~\cite{Goebel2021}, as long as the texture can be described by a set of collective coordinates, where the only term that needs to be recalculated is $V^i_{nn^\prime}$.

\textit{From stochastic equation of motion to massive dynamics.}---In order to extract skyrmion mass from $\eta_{ij}$, we first separate \cite{Hurst2020} the real (i.e., spectral) from the imaginary (i.e., dissipative) component \mbox{$\eta_{ij}(\omega) = [J_{ij}(\omega) + i f_{ij}(\omega)]/\omega$}, where we use the time-translation invariance $\check G_n(t,t^\prime) = \check G_n(t-t^\prime)$ in steady-state nonequilibrium to Fourier transform $t-t^\prime$ into frequency $\omega$ space. These two components are then given by
\begin{subequations}\label{eq:spectralDissipative}
\begin{eqnarray}
J_{i j} &=& \frac{\pi}{2} \sum_{nn^\prime} V^i_{nn^\prime} V^j_{n^\prime n}\left(h(\epsilon_{n^\prime})-h(\epsilon_n) \right)(\epsilon_{n^\prime}-\epsilon_n)^2 \nonumber \\
&\times& \delta\left(\omega-\epsilon_{n^\prime}+\epsilon_n\right) \\
f_{i j} &=& \frac{\omega^2}{2} \sum_{nn^\prime} V^i_{nn^\prime} V^j_{n^\prime n} \left(\frac{h(\epsilon_{n^\prime})-h(\epsilon_n)}{\omega-\epsilon_{n^\prime}+\epsilon_n}\right), 
\end{eqnarray}
\end{subequations}
where $h(\epsilon_n) = \tanh((\epsilon_n - \mu)/2k_B T) = 1-2f(\epsilon_n)$ is related to the Fermi function $f(\epsilon_n)$ at chemical potential $\mu$ and temperature $T$. The matrix elements $V^i_{nn^\prime}$ can only be calculated numerically by discretizing 2D space to become a square lattice of spacing $a$ (with open boundary conditions), length $L$, width $W$, and the electronic hopping parameter $\gamma = \hbar^2/2m_ea^2$. The skyrmion texture was approximated using the standard circular DW ansatz \cite{romming2015, Siemens2016, Xing2022} in polar coordinates $(\rho, \phi)$
\begin{subequations}
\begin{eqnarray}
    M^x_0(\rho, \phi) &=& \cos(\phi + \varphi){\rm sech}\left(\frac{\rho-\rho_{\rm sky}}{\Delta}\right), \\
    M^y_0(\rho, \phi) &=& \sin(\phi + \varphi){\rm sech}\left(\frac{\rho-\rho_{\rm sky}}{\Delta}\right), \\
    M^z_0(\rho, \phi) &=& \tanh\left(\frac{\rho-\rho_{\rm sky}}{\Delta}\right),
\end{eqnarray}
\end{subequations}
where $\rho_{\rm sky}$ is the skyrmion radius, $\Delta$ is the DW width between the inside and outside of the skyrmion, $\varphi$ is the helicity, and the subscript 0 emphasizes that this is a rigid solution. 

We find that not all components of $\eta_{ij}(\omega)$ are equally relevant [Fig. \ref{figMass}(a)]. That is, at low frequencies the diagonal dissipative part, $f_{ii}=\tilde m_i\omega^2$, dominates over the spectral diagonal $J_{ii}$ and the off-diagonal terms $f_{ij}$ and $J_{ij} (i\neq j)$. Transforming Eq.~\eqref{eqRawEOM} back into the time domain we obtain
\begin{equation}
    m_i\ddot R^i + \epsilon_{ij}\tilde Q \dot R^j = F^i + \xi^i,
\end{equation}
where the mass $m_i$ \textit{emerges from the quadratic dependence of} $f_{ii}$. Due to the finite-size of numerical calculations to extract the mass, we eliminate its artifactual dependence on the area of the lattice $LW$ [Fig.~\ref{figMass}(b)] by using $m_i = \tilde m_i/LW$, where $m_i$ is the true mass.

In the regime in which conduction electrons are spin-unpolarized, $\mu \gg J_{sd}$, due to equal number of spin-up and spin-down conduction electrons, only $J_{sd}$ and the width of the nanowire $W$ affect value of skyrmion mass [Figs.~\ref{figMass}(c) and \ref{figMass}(d)]. Therefore, we set $Q=1$, $\varphi = 0$, $\rho_{\rm sky}=5a$, and $\Delta=1a$ since these intrinsic skyrmion properties are irrelevant to its electron-induced mass. This supports Ref.~\cite{Martinez2017}, where the relevance of skyrmion mass due to conduction electrons was conjectured, but its form was introduced phenomenologically into the LLG equations without any dependence on the topological charge $Q$. 

Figure~\ref{figMass}(c) demonstrates that $m_i$ decreases with increasing $J_{sd}$. Since, in general, misalignment between $\mathbf{M}(\mathbf r,t)$ and electronic spin density $\langle \hat{\mathbf{s}}(\mathbf r,t)\rangle$ decreases with $J_{sd}$ (for large enough $J_{sd}$~\cite{Petrovic2021}), the property of mass vs. $J_{sd}$ in Fig.~\ref{figMass}(c) originates from STT~\cite{Tatara2019, Petrovic2021, Bajpai2020} $\propto \langle \hat{\mathbf{s}}\rangle (\mathbf{r}) \times \mathbf{M}(\mathbf{r})$.

The other parameters affecting the mass of the skyrmion have to do with the geometry of the area through which skyrmion propagates. In the case of square geometry $L=W$ of sufficiently large size, the mass does not depend on the size [Fig.~\ref{figMass}(d)]. However, as we increase transverse confinement $L\gg W$, thereby forcing the skyrmion to propagate through a quasi-one-dimensional nanowire as often employed experimentally~\cite{Parkin2015, Yu2017}, we find different masses for different directions of propagation [Fig.~\ref{figMass}(d)]. In particular, the mass of skyrmion $m_y$ for propagation along the transverse direction is larger, $m_y > m_x$, than the mass of skyrmion propagating longitudinally along the nanowire  [Fig.~\ref{figMass}(d)]. Thus found \textit{anisotropy} of electron-induced skyrmion mass is a new way to manipulate its dynamics. Similar difference between skyrmion masses in square vs. nanowire geometry was observed when the mass is generated by interaction with magnons~\cite{Psaroudaki2018}.

\begin{figure}
    \centering
    \includegraphics[width=\linewidth]{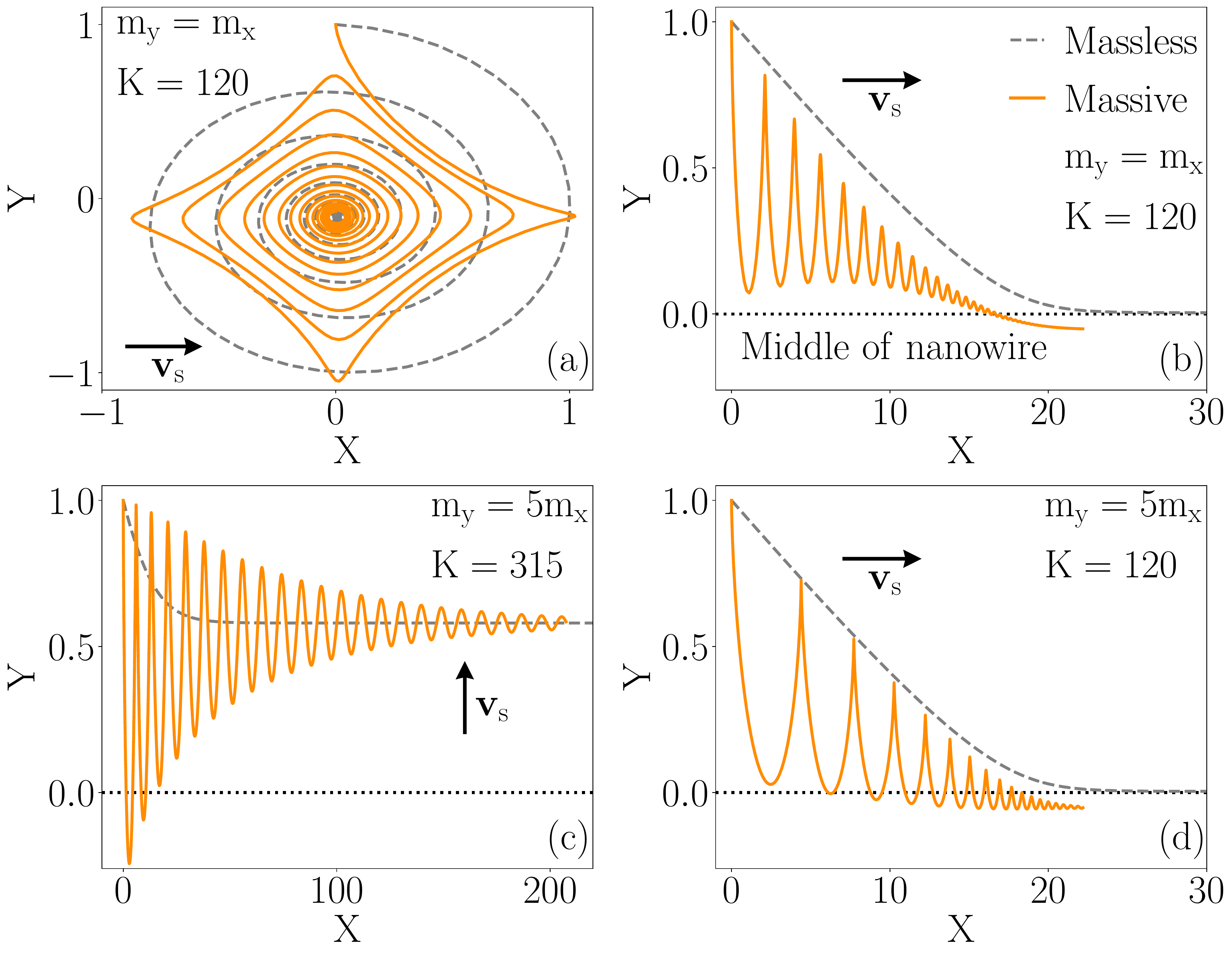}
    \caption{(a) Trajectory of the skyrmion center $\mathbf{R}_c\equiv\mathbf{R}=(X,Y)$ in a 2D harmonic trap obtained by solving Eq.~\eqref{eqFinalEOM}. (b)--(d) The same trajectories but with the harmonic potential applied only along the $y$-axis to simulate nanowire geometry. Massive (massless) trajectories are plotted by orange (dashed grey) line. Note that orange line in panel (a) is a hypocycloid. Oscillations in the trajectories are due to the mass, whose properties are studied in Fig.~\ref{figMass} and whose magnitude and \textit{anisotropy} affect the frequency of such oscillations in the orange trajectories. The vector $\mathbf{v}_s$ denotes the direction of injected electronic charge current in Eq.~\eqref{eqFinalEOM} and $K$ measures the strength of the harmonic confining potential.}
    \label{figTrajectories}
\end{figure}

In order to simulate current-driven dynamics of massive skyrmions, we integrate Eq.~\eqref{eqFinalEOM} with $D=5.5\pi$, $\beta=0.025$, and $\alpha=0.05$ chosen to be the same as in Ref.~\cite{Martinez2017}). For simplicity, we suppress~\cite{Hurst2020} stochastic effects on dynamics due to quantum noise by setting $\xi_i = 0$. In the case where the skyrmion is in a 2D harmonic trap $U=KR^2/2$, resembling our numerical calculation in Fig.~\ref{figMass} for $L=W$, the mass is isotropic. By ignoring damping, $D=0$, solutions of Eq.~\eqref{eqFinalEOM} can be expressed in terms of four normal modes, $\Xi^\pm_\circlearrowleft =  (\cos\omega^\pm t, \sin\omega^\pm t)$ and $\Xi^\pm_\circlearrowright = (-\cos\omega^\pm t, \sin\omega^\pm t)$, with frequencies $\omega^\pm = \big(\tilde Q \pm \sqrt{\tilde Q^2 + 4mK}\big)/2m$, where the subscript denotes clockwise or counter-clockwise rotation of the normal mode. Therefore, the skyrmion center traces hypocycloids, whose amplitude will decrease if damping is non-zero $D>0$, as illustrated in Fig.~\ref{figTrajectories}(a). For easy comparison, the corresponding trajectory of a massless skyrmion [$m=0$ in Eq.~\eqref{eqFinalEOM}] is plotted as a dashed grey line in Fig.~\ref{figTrajectories}.

Alternatively, in a nanowire the skyrmion is only confined along the $y$-axis, which we simulate by using $U=Ky^2/2$ in Eq.~\eqref{eqFinalEOM} and plot thus computed trajectories in Fig. \ref{figTrajectories}(b)--(d). When electronic current flows along the longitudinal nanowire direction, the skyrmion drifts [Figs.~\ref{figTrajectories}(b) and \ref{figTrajectories}(d)] towards the middle of the nanowire due to damping and then continues following the direction of the current. In this case, the skyrmion mass causes oscillations in the trajectory, while its \textit{anisotropy} affects the frequency of such oscillations, as seen by contrasting Figs.~\ref{figTrajectories}(b) and \ref{figTrajectories}(d). In the case of transversely injected current [Fig. \ref{figTrajectories}(c)], skyrmion still moves longitudinally but with its center being away from the middle of the nanowire. In this case as well, the mass induces oscillations in the trajectory [Fig.~\ref{figTrajectories}(c)].

\textit{Conclusions.}---In addition to previously discussed origins of skyrmion mass in the Thiele Eq.~\eqref{eqFinalEOM}---such as due to its nonrigidity~\cite{Makhfudz2012, Wu2022}, magnons~\cite{Psaroudaki2021}, phonons~\cite{Capic2020} and geometry~\cite{Pavlis2020}---our Schwinger-Keldysh field-theoretical approach reveals how interaction with spins of conduction electrons leads to skyrmion mass, even when its shape remains rigid. Such mass becomes \textit{anisotropic}, i.e., dependent on the direction of skyrmion center motion, in nanowire geometries with strong transverse confinement. This, in turn, allows for massive current-driven dynamics, as studied phenomenologically in Ref.~\cite{Martinez2017}, which is not possible with widely-discussed skyrmion mass due to its nonrigidity~\cite{Wu2022}. Although our numerical evaluation of electronically-induced skyrmion mass does not allow us to determine its precise value due to finite-size effects, we estimate it to be in the range \mbox{$10^{-31}$--$10^{-25}$ kg}. These numbers would make the electronic contribution to skyrmion mass comparable or greater than contributions from other mechanisms such as magnons~\cite{Psaroudaki2018}. Furthermore, we predict a clear signature of electronically-induced skyrmion mass---oscillations superimposed on translational motion of skyrmion center with their frequency depending on the mass---which could be tested through experiments or numerically exact quantum-classical simulations of electron-skyrmion systems~\cite{Petrovic2018,Petrovic2021,Bostroem2019}.

This work was supported by the US National Science Foundation (NSF) Grant No. ECCS 1922689.

\bibliography{biblio}

\end{document}